# A perturbation theory for water with an associating reference fluid

Bennett D. Marshall

*ExxonMobil Research and Engineering, 22777 Springwoods Village Parkway, Spring TX 77389 USA*

**Abstract**

The theoretical description of the thermodynamics of water is challenged by the structural transition towards tetrahedral symmetry at ambient conditions. As perturbation theories typically assume a spherically symmetric reference fluid, they are incapable of accurately describing the liquid properties of water at ambient conditions. In this paper we address this problem by introducing the concept of an associated reference perturbation theory (APT). In APT we treat the reference fluid as an associating hard sphere fluid which transitions to tetrahedral symmetry in the fully hydrogen bonded limit. We calculate this transition in a theoretically self-consistent manner without appealing to molecular simulations. This associated reference provides the reference fluid for a second order Barker-Hendersen perturbative treatment of the long-range attractions. We demonstrate that this new approach gives a significantly improved description of water as compared to standard perturbation theories.

Bennettd1980@gmail.com

# I: Introduction

The theoretical description of the thermodynamics of water has been the focus of intense investigation over the last several decades.[1] The structural transition seen in water at ambient and super cooled conditions is due to the tetrahedral symmetry of fully hydrogen bonded water. This results in anomalous thermodynamic behavior such as minima in the isothermal compressibility and isobaric heat capacity, as well as maximum in the liquid density as a function of temperature.[2] Molecular simulations are often able to predict these anomalous features of water using relatively simple classical water models[3,4]. There has also been progress made in terms of the description of water in terms of quasi-chemical theory paired with molecular simulation.[5,6]

Theoretical approaches such as the single bond approach of Truskett et al.[7] and the cluster theory of Dahl and Andersen[8] qualitatively reproduce the anomalous properties of liquid water. However, these approaches lack quantitative accuracy and have not been extended to allow for the description of multi-component mixtures.

While the theoretical description of pure water is a worthy academic pursuit, nearly all thermodynamics problems of practical interest involve multi-component mixtures. Equations of state based on perturbation theory[9,10] are widely employed in academia and industry to describe the thermodynamics and phase behavior of fluid mixtures.[11–14] Perturbation theories separate the intermolecular potential into a short ranged rapidly varying contribution, and a longer range slowly varying contribution. It is the short range contribution which determines the structure of the fluid[15,16], with the long range contribution acting as the perturbation.

In general, successful application of perturbation theory requires that the attractions which are perturbing the reference fluid structure do not significantly change the fluid structure. In most applications of perturbation theory to simple and associating fluids, a spherically symmetric repulsive reference state is chosen.[10,17,18] For many fluids, this provides a proper reference state. However, for water at ambient conditions it does not. This is because hydrogen bonding significantly changes the structure of water resulting in tetrahedral symmetry when all water molecules are fully hydrogen bonded.

Remsing *et al.*[16] split the Columbic contribution of the SPC/E[19] potential of water into a short range and long range contributions using local molecular field theory[20]. It was demonstrated using molecular dynamics simulation, that when the short ranged Columbic forces where explicitly treated in the simulation, and the long range contributions were treated as a perturbation, that the truncated SPC/E potential gave an excellent representation of the full SPC/E potential. That is, if hydrogen bonding is included in the reference system, perturbation theory can be accurately applied to water.

In this work we take the approach of including hydrogen bonding in the reference fluid to develop a new perturbation theory for water. However, we wish to develop a purely theoretical approach which will be extendable to multi-component mixtures. The reference fluid is taken to be a fluid of hard spheres with 4 (2 donor and 2 acceptor) association sites. The association contribution of the reference free energy is obtained using Wertheim's[18,21,22] first order thermodynamic perturbation theory (TPT1). The long ranged attractions are treated as a square well attraction in Barker-Hendersen second order perturbation theory (BH2).[9]

Application of BH2 requires knowledge of the integral of the reference system correlation function. In this work, we exploit the tetrahedral symmetry of fully hydrogen bonded water to develop a simple and accurate representation of this integral as a function of the degree of hydrogen bonding. We call this new approach associated reference perturbation theory (APT).

It is demonstrated that APT gives a significantly improved equation of state as compared to theories which employ BH2 with a hard sphere reference fluid.

## II: Theory

In this section, we develop a new perturbation theory which accounts for the tetrahedral geometry of liquid water. Water is taken to be a single sphere of diameter $d$ with 4 hydrogen bonding sites: two hydrogen bond acceptor sites ($O_1$, $O_2$) and two hydrogen bond donor sites ($H_1$, $H_2$) in the overall set $\Gamma = \{O_1, O_2, H_1, H_2\}$. The overall potential of interaction between two water molecules is given as

$$\varphi(12) = \varphi_{hs}(r_{12}) + \varphi_{sw}(r_{12}) + \varphi_{as}(12) \tag{1}$$

Where (1) represents the position $\mathbf{r_1}$ and orientation $\mathbf{\Omega_1}$ of molecule 1, $\varphi_{hs}$ is the potential of the hard sphere reference fluid, $\varphi_{sw}$ is an attractive perturbation due to an isotropic square well attraction of depth $\varepsilon$ and range $\lambda$. Lastly, the association contribution to the intermolecular potential is taken as the sum over site-site potentials[23]

$$\varphi_{as}(12) = \sum_{A \in \Gamma} \sum_{B \in \Gamma} \varphi_{AB}(12) \tag{2}$$

For the site-site potential we assume conical square well association sites[18]

$$\varphi_{AB}(12) = \begin{cases} -\varepsilon_{AB} & r_{12} \leq r_c \text{ and } \theta_{A1} \leq \theta_c \text{ and } \theta_{B2} \leq \theta_c \\ 0 & otherwise \end{cases} \tag{3}$$

Where $r_c$ is the maximum distance between molecules for which association can occur, $\theta_{A1}$ is the angle between the center of site $A$ on molecule 1 and the vector connecting the two centers, and $\theta_c$ is the maximum angle for which association can occur. With this, if two molecules are both positioned and oriented correctly, a bond is formed and the energy of the system is decreased by a factor $\varepsilon_{AB}$. A diagram of two molecules interacting with conical square well association sites can be found in Fig. 1.

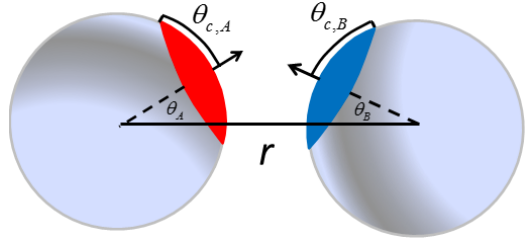

**Figure 1 (color online):** Diagram of interacting conical square well association sites

It is common practice in the development of perturbation theories for associating fluids to consider non-association attractions in the reference system for the association perturbation theory. For instance, in the case of associating Lennard-Jones spheres, the Lennard-Jones fluid is taken as the reference fluid for the perturbation theory.[24] Here we take a different approach. We consider the hard sphere + association potential to be the reference system for the square well attraction. The Helmholtz free energy is then given by

$$A = A_{ref} + A_{sw} \tag{4}$$

where $A_{ref}$ is free energy of the associating reference system and $A_{sw}$ is the change in free due to isotropic square well attraction. The reference free energy is composed of ideal gas, hard sphere repulsions, and an association perturbation

$$A_{ref} = Nk_bT(\ln \rho \Lambda - 1) + A_{hs} + A_{as} \tag{5}$$

$\Lambda$ gives the temperature dependent ideal gas contributions and $A_{hs}$ is the excess free energy due to hard sphere repulsion given by the Carnahan and Starling equation of state[25].

The association contribution to the free energy $A_{as}$ is evaluated using Wertheim's first order perturbation however, TPT1 alone does not account for the structural transition to tetrahedral symmetry. For more info on the TPT1 approach see supplementary material.

With the reference fluid now defined, we treat the square-well attractions in second order Barker-Hendersen[9,26] perturbation theory (BH2) as a perturbation to the associating reference fluid

$$A_{sw} = \frac{\varepsilon}{k_b T} A_1 + \left(\frac{\varepsilon}{k_b T}\right)^2 A_2 \qquad (6)$$

where $A_1$ and $A_2$ are the first and second order perturbation free energies and $A_2$ is evaluated using the local compressibility approximation. Evaluation of $A_1$ and $A_2$ in Eq. (6) requires knowledge of the integral of the reference system pair correlation function $g_{ref}$

$$I_{ref}(a) = \int_1^{a/d} x^2 g_{ref}(x) dx \qquad x = r/d \qquad (7)$$

The integral $I_{ref}(a)$ is related to the reference system coordination number $N_{ref}$ through the following relation

$$N_{ref}(a) = 4\pi \rho d^3 I_{ref}(a) \qquad (8)$$

Hence $I_{ref}$ controls the number of molecules for which a given molecule will share isotropic square well attractions. The standard approach is to assume a hard sphere reference fluid

$$I_{ref} = I_{hs} \qquad (9)$$

While this is often a very good approximation for non-associating simple fluids, it will fail for water at ambient conditions where water is highly structured with near tetrahedral symmetry. That is, at ambient conditions Eq. (9) will predict a coordination number which is too large. This will result in the over-prediction of the isotropic square well attractions. Hence, we can consider Eq. (9) to be appropriate as a limiting high temperature case.

Another limit would be the case of pure tetrahedral symmetry $N_{ref}(r_c) = 4$. For the tetrahedral limit we can employ Eqn. (8) to solve for $I_{ref}$ up to the critical radius $r_c$ analytically as

$$I_{tet}(r_c) = \int_{1}^{r_c/d} x^2 g_{tet}(x)dx = \frac{1}{\pi \rho d^3} \tag{10}$$

In Eq. (10) we have enforced that if all water molecules are fully bonded there will be tetrahedral coordination within the shell $r_c$. The correlation function $g_{tet}$ is the pair correlation function between two water molecules in a fully hydrogen bonded fluid. Defining $I_{tet}$ over the full square well range $\lambda$ for a fully hydrogen bonded fluid

$$I_{tet}(\lambda) = \int_{1}^{r_c/d} x^2 g_{tet}(x)dx + \int_{r_c/d}^{\lambda/d} x^2 g_{hs}(x)dx = \frac{1}{\pi \rho d^3} + I_{hs}(\lambda) - I_{hs}(r_c) \tag{11}$$

In equation (11) we have assumed, for a fully hydrogen bonded system, that within the range $r_c$ the reference fluid is that of tetrahedral coordination, while in the range $r_c \leq r \leq \lambda$ the hard sphere reference fluid is used. The integral $I_{hs}$ is evaluated using the real function solution of the Ornstein-Zernike equation within the Percus-Yevick approximation of Chang and Sandler[26,27].

Equations (7) - (8) are developed by considering a molecule (labelled 0) at the origin ($r = 0$). For the 4-site associating reference fluid, molecule 0 could be hydrogen bonded $k = 0 - 4$ times. For the case that molecule 0 is bonded $k$ times, the number of molecules within a shell $a$ will be $N_k(a) = 4\pi d^3 \rho I_k(a)$, where $I_k$ is the reference fluid integral for the case that molecule 0 is bonded $k$ times. For the hydrogen bonding reference fluid we then write $N_{ref}$ as an average over all possible bonding states

$$N_{ref}(a) = 4\pi \rho d^3 \sum_{k=0}^{4} \chi_k I_k(a) \tag{12}$$

where $\chi_k$ is the fraction of molecules bonded $k$ times. Comparing Eqns. (8) and (12)

$$I_{ref}(a) = \sum_{k=0}^{4} \chi_k I_k(a) \tag{13}$$

In general, the $I_k$ will depend on the hydrogen bonded state of the system; however, Overduin and Patey[28] demonstrated that water molecules tend to cluster based on the tetrahedral order parameter. Hence, it is more likely that if molecule 0 is fully hydrogen bonded, it will be surrounded by other fully hydrogen bonded water molecules than would be the case if molecule 0 were less than fully hydrogen bonded. On this basis, we assume $I_4 = I_{tet}$. Similarly, if molecule 0 is less than fully hydrogen bonded, it is more likely it will be surrounded by other molecules which are not fully hydrogen bonded. As the anomalous properties of water are the result of tetrahedral symmetry[2], for these integrals we revert to the standard perturbation treatment $I_k = I_{hs}$ for $k < 4$. Combining Eqns. (11) – (13) with $I_4 = I_{tet}$ and $I_k = I_{hs}$ for $k < 4$ we obtain the final result

$$I_{ref}(\lambda) = (1-\chi_4)I_{hs}(\lambda) + \chi_4 I_{tet}(\lambda) = I_{hs}(\lambda) + \chi_4\left(\frac{1}{\pi\rho d^3} - I_{hs}(r_c)\right) \tag{14}$$

Equation (14) completes the associated reference perturbation theory for water. We have incorporated t parameters of the theory are the diameter $d$, association parameters $\varepsilon_{AB}$, $\theta_c$, $r_c$ and isotropic square well parameters $\varepsilon$ and $\lambda$. We discuss evaluation of the parameters in section III.

**III: Parameterization and model results**

In this section we develop parameters for the theory developed in section II. We consider two cases, in the first case we employ the associated reference perturbation theory (APT) with $I_{ref}$ given by Eq. (14), while for the second case the theory is treated as a hard sphere reference perturbation theory (HSPT) with $I_{ref}$ given by Eq. (9). For each case, the parameters include the hard sphere diameter $d$, isotropic square well depth $\varepsilon$, square well range $\lambda$, hydrogen bond energy $\varepsilon_{AB}$, critical angle for hydrogen bonding $\theta_c$, as well as well the hydrogen binding range $r_c$. The parameters are obtained by fitting the model to saturated vapor pressure and liquid density data in the range 273.15 K < T < 580 K.

Table 1 gives the regressed parameters for both APT and HSPT. The diameters $d$ obtained for both models are consistent with neutron diffraction[29] data which shows that the first maximum in oxygen-oxygen correlation function in liquid water is located at a distance of $d$ = 2.75 Å. This same neutron diffraction data shows the first minima in the oxygen-oxygen correlation function at 3.5 Å. Soper et al.[29] took this minimum to represent the maximum hydrogen bonding distance giving $r_c / d$ = 3.5 / 2.75 = 1.27$d$. This value is consistent with the regressed values in Table 1. Finally, Luck[30] estimated (from spectroscopic data) the energy of a liquid phase water-water hydrogen bond to be 1862$k_b$ which is consistent with the regressed results in Table 1.

Table 2 gives average absolute deviations (AAD) between model and experimental data. For comparison we include perturbed chain statistical associating fluid theory (PC-SAFT[31]) results for a standard PC-SAFT model fit to the same data (method I of ref[32]) . As can be seen, both APT and HSPT significantly outperform PC-SAFT. The relatively poor performance of PC-SAFT for water is the result of the PC-SAFT contribution for isotropic attractions. PC-SAFT is also based on BH2 with a TPT1 treatment of hydrogen bonding. However, a significant level of empiricism[13] was imposed on the BH2 model within PC-SAFT to develop "universal constants". These universal constants were fit by considering phase behavior of $n$-alkanes and square well chains. As the structure of water is significantly different than the training set used in the model development, it is not optimal for use as an equation of state for water. On the other hand (forgetting APT for a moment), using BH2 with an analytical representation[26] of the integral $I_{hs}$ gives a purely theoretical basis, and overall much better performance for water.

Comparing APT and HSPT, APT gives a significant improvement in liquid densities; however, for this choice of parameters HSPT gives a slightly better representation of the vapor pressures.

| Method | $d$(Å) | $\varepsilon/k_b$ | $\lambda/d$ | $\varepsilon_{AB}/k_b$ (K) | $\theta_c$ | $r_c/d$ |
|---|---|---|---|---|---|---|
| APT | 2.8440 | 241.489 | 1.8894 | 1891.836 | 28° | 1.2939 |
| HSPT | 2.8200 | 239.246 | 1.6037 | 1868.186 | 41.765° | 1.2458 |

**Table 1:** Model parameters

| Method | AAD% $P_{sat}$ | AAD% $\rho_L$ |
|---|---|---|
| APT | 0.82 | 0.30 |
| HSPT | 0.5 | 1.52 |
| PC-SAFT[32] | 2.3 | 4.1 |

**Table 2:** Average absolute deviations for three equations of state with vapor pressure ($P_{sat}$) and liquid density ($\rho_L$) in the temperature range 273.15 K < T < 580 K

Figure 2 compares model predictions to experimental vapor pressures for the APT model (HSPT model results are visually indistinguishable from APT). Figure 3 shows the phase diagram as predicted by APT (left panel) as well as a comparison of both APT and HSPT for the prediction of saturated liquid densities near ambient temperatures (right panel). As can be seen,

APT significantly outperforms HSPT for the prediction of liquid densities. In fact, APT does an outstanding job predicting the liquid densities at ambient conditions. However, for this choice of parameters, APT does not produce a density maximum. It is possible to obtain the density maxima using APT, however this results in a loss of accuracy in the representation of vapor pressures. As is common place in perturbation theories, both approaches predict too high of a critical temperature.

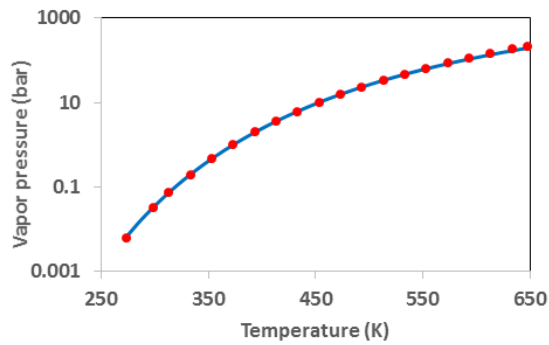

**Figure 2 (color online):** Comparison of APT vapor pressure model results (curve) to experimental data[33]

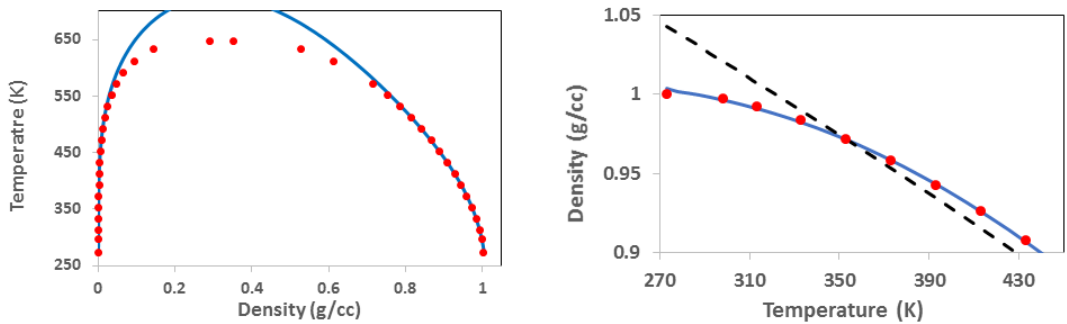

**Figure 3 (color online):** Left: $T$-$\rho$ phase diagram for water. Curve gives APT model results and symbols give experiment[33]. Right: Zoomed in on saturated liquid densities neat ambient conditions. Solid curve gives APT results and dashed curve gives HSPT results.

The improvement in liquid density predictions of the APT is a result of the fact that the structural change in the fluid which results from hydrogen bonded is included in $I_{ref}$ via Eq. (14).

This can be seen in Fig. 4 which plots the reference system coordination number Eq. (8) within a shell of 3.3Å for a saturated liquid using both APT and HSPT. We have also included molecular simulation results[34,35] for the coordination number of saturated liquid water in the shell of 3.3Å using both TIP4P/2005[3] and iAMOEBA[4] force fields. As can be seen, APT allows for the self-consistent calculation of a more appropriate reference fluid than is possible using HSPT. $N_{ref}$ calculated using HSPT is linear in temperature. The non-linearity of the APT coordination number is the result of the increase in fraction of molecules bonded 4 times; this is shown in the right panel of Fig. 4.

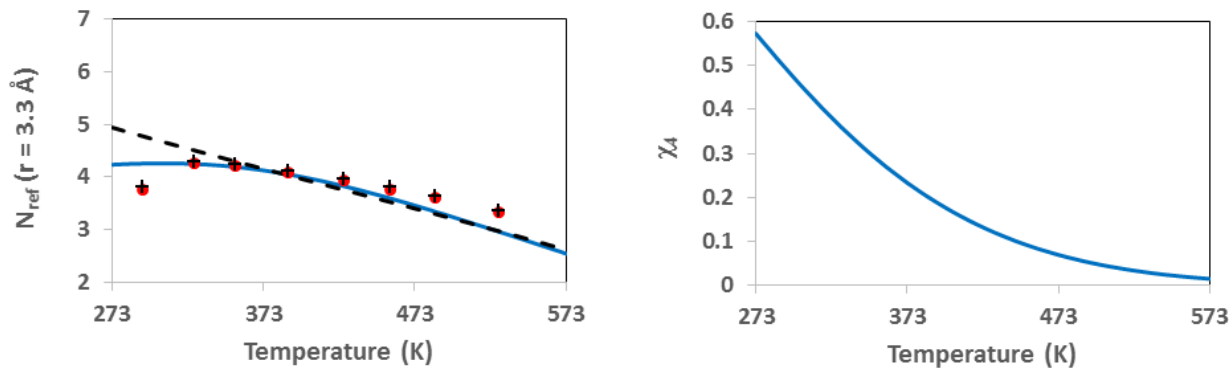

**Figure 4 (color online):** Left: Reference system coordination number at for a saturated liquid in a shell of 3.3 Å. Solid curve gives APT and dashed curve gives HSPT. Symbols give molecular simulation results[34,35] using iAMOEBA[4] (circles) and TIP4P/2005[3] (crosses). Right: Fraction of water molecules bonded 4 times for a saturated liquid predicted from APT.

Figure 5 gives results for the volume expansivity $\alpha = -\partial \ln\rho/\partial T$ at atmospheric pressure. As can be seen, APT gives a much better representation of the data than HSPT. In fact, for temperatures $T > 310$ K APT is nearly quantitative. However, as noted on the phase diagram in Fig. 3, APT does not produce the density maximum ($\alpha < 0$). APT can be parameterized to reproduce the density maximum; however, this would come at the expense of accuracy in the representation of vapor pressure.

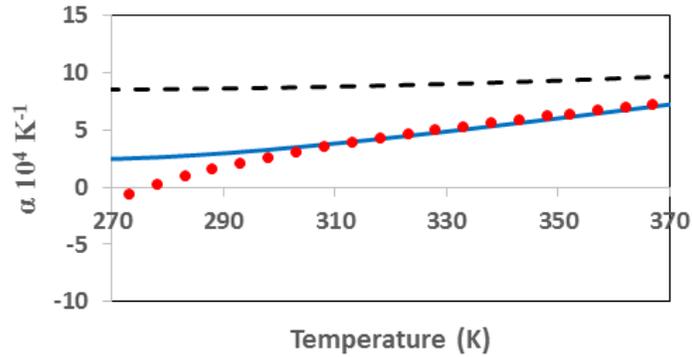

**Figure 5 (color online):** Volume expansivity of water versus $T$ at atmospheric pressure. Curves have same meaning as Fig. 4 and symbols give experimental data[36]

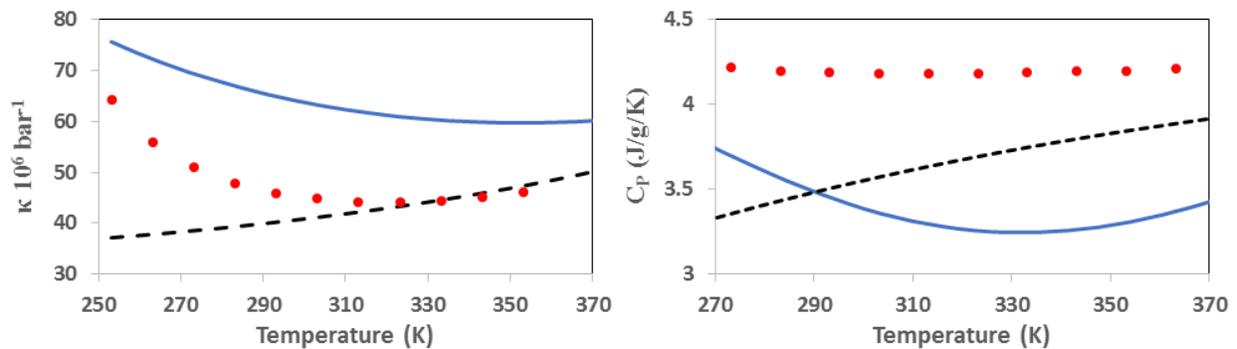

**Figure 6 (color online):** Left: Isothermal compressibility versus $T$ at a pressure of 1 bar. Lines have same meaning as Fig. 4 and symbols give data[37]. Right: Isobaric heat capacity at 1 bar. Symbols give experimental data.[38]

In Fig. 6 we compare theoretical predictions and experiment for the second derivative properties isothermal compressibility κ and isobaric heat capacity $C_p$ of liquid water at 1 bar. While neither APT nor HSPT is quantitatively accurate for these properties, APT is qualitatively consistent with experiment by exhibiting the anomalous minima as a function of $T$. Table 3 compares APT and experiment for the temperature at which the minima occur.

|     | $T_{min}$(Data) | $T_{min}$(APT) |
|-----|-----------------|----------------|
| κ   | 319 K           | 342 K          |
| $C_P$ (J/g/K) | 308 K | 334 K          |

**Table 3:** Comparison of experimental and APT predicted temperatures at which minima occur in the isothermal compressibility and isobaric heat capacity

**IV: Conclusions**

We have developed a new perturbation theory for water which treats an associating hard sphere fluid as a reference fluid for long ranged attractions through BH2. The transition of the BH2 reference fluid to tetrahedral symmetry in the fully hydrogen bonded limit is calculated self-consistently in the theory. It was demonstrated that the new approach gave substantially improved liquid state predictions as compared to the hard sphere reference case, as well as the popular PC-SAFT equation of state.

A benefit of this purely theoretical approach is that it is amenable to extension to multi-component mixtures. TPT1 is general for multi-component mixtures[11] and BH2 can be extended to mixtures using appropriate mixing rules.[13,39] This will be the subject of a future publication.

**Acknowledgments:**

The author thanks Dilip Asthagiri for sharing unpublished simulation data on the coordination number of saturated liquid water in Fig. 4.